\documentclass[amsmath,amssymb]{revtex4}
  \usepackage{graphicx}

\begin{document}
\title{Brownian oscillator with time-dependent strength:   A delta function 
protocol}
\author{Alex V. Plyukhin}
\email{aplyukhin@anselm.edu}
 \affiliation{ Department of Mathematics,
Saint Anselm College, Manchester, New Hampshire 03102, USA 
}

\date{\today}

\begin{abstract}
We consider a classical Brownian oscillator of mass $m$ driven from an arbitrary initial state 
by varying the stiffness $k(t)$ of the harmonic potential according to the protocol 
$k(t)=k_0+a\,\delta(t)$, involving the Dirac delta function. 
The microscopic work performed on the oscillator is shown to be
$W=(a^2/2m)\,q^2-a q v$, where  $q$ and $v$  are the coordinate and velocity
in the initial state. 
If the initial distribution of $q$ and $v$ is the equilibrium
one with temperature $T$, the average work is
$\langle W \rangle=a^2T/(2m\,k_0)$ and  
the distribution $f(W)$ has the form of the product of exponential and modified Bessel functions. The distribution is asymmetric and diverges as $W\to 0$.
The system's  response for $t>0$ is evaluated for specific models.
\end{abstract}


\maketitle

\section{Introduction}
Brownian motion in a time dependent 
harmonic potential is a stochastic process important from both 
theoretical and experimental points of view. It provides an insightful example and testing ground for subtle theoretical results,  like fluctuation theorems~\cite{PP}. 
It is also directly relevant to important experimental techniques
to study mesoscopic systems using optical traps. In
such experiments, the system is driven from thermal equilibrium
by varying  the strength of the trap, i.e. the stiffness of the harmonic potential   $k(t)$, according to a
certain protocol,  
and the work $W$ on the system, as well as the system's response, are recorded~\cite{traps1,traps2}.

The optical trap setup  has been used also 
to design Brownian engines with a single optically trapped Brownian particle 
as a working substance and with the trap stiffness $k(t)$ as a control parameter~\cite{Blickle,Martinez,SM,HR,Ryabov_review}. Such machines often operate out of equilibrium
using driving protocols of  finite  duration and thus belong to the realm of finite-time thermodynamics~\cite{FT,Watanabe}.

The steps of isothermal compression (expansion) of the macroscopic Carnot cycle correspond in Brownian engines to the increase (decrease) of the stiffness $k(t)$ at fixed temperature during a finite switching time interval $(0,t_s)$. 
The corresponding work $W$ strongly fluctuates and is governed by equations of stochastic dynamics of Langevin or Fokker-Planck types with the time-dependent potential $V(q,t)=k(t)q^2/2$. 
The microscopic work can be written as~\cite{Jar}
\begin{eqnarray}
W=\int_0^{t_s} \frac{dV(q,t)}{dk}\, \dot k(t)\,dt=
    \frac{1}{2}\,\int_0^{t_s}  q^2(t)\,\dot k(t)\, dt,
    \label{W_general0}
\end{eqnarray}
and the average work is a functional of the mean-square displacement
$\langle q^2(t)\rangle$, 
\begin{eqnarray}
\langle W\rangle=
    \frac{1}{2}\,\int_0^{t_s}  \langle q^2(t)\rangle\,\dot k(t)\, dt. 
    \label{work_general}
\end{eqnarray}
Since the microscopic work is a nonlinear functional of 
$q$, the work distribution is not  Gaussian  in general.
For overdamped Langevin dynamics with white Gaussian noise the asymptotic distribution for large absolute values of work was found to have the form
\begin{eqnarray}
    f(W)= c_1\frac{e^{-c_2|W|}}{\sqrt{|W|}},
    \label{w_dist_asymp}
\end{eqnarray}
where $c_1, c_2$ depend on the specific form of the protocol $k(t)$~\cite{Engel,NE}. This form, with 
an exponential tail  and a power-law prefactor,
was found for $f(W)$ also for Langevin dynamics beyond the overdamped limit~\cite{Kwon}. 
On the other hand, the central part of the work distribution  was predicted to have 
an approximately  Gaussian form
in the limit of a slow driving~\cite{Seifert,Speck}.

The  evaluation of the work distribution 
in closed form for an arbitrary protocol
and for the whole range of $W$
appears to be a rather formidable task. 
However, for   
a special protocol, often referred to as the instantaneous quench~\cite{PP}, the calculations are very simple. In that protocol  at $t=0$
the stiffness is instantaneously switched  from $k_0$ to a new value $k_1$, i.e. 
\begin{eqnarray}
    k(t)=k_0+(k_1-k_0)\,\theta(t),
    \label{k_quench}
\end{eqnarray}
where $\theta(t)$ is the step function.
The corresponding microscopic work {\it on the system} 
is equal to the difference of the 
system's potential energy 
\begin{eqnarray}
    W=\frac{k_1-k_0}{2}\,q^2,
    \label{W_aux}
\end{eqnarray}
where $q=q(0)$. Because the protocol's duration is effectively zero, 
statistics of $W$ does not depend on details of stochastic dynamics (such as statistics of the noise) and is completely determined by 
the distribution of the oscillator's initial coordinate $q$. Suppose at $t<0$ the oscillator with the stiffness $k=k_0$ is in thermal equilibrium and the 
coordinate and velocity are distributed according to the canonical distribution
\begin{eqnarray}
    \rho_0(q,v)=\frac{1}{Z}\, \exp\left(-\frac{m\,v^2}{2T}-\frac{k_0\,q^2}{2T}\right). 
    \label{dist}
\end{eqnarray}
Here and below  temperature is in energy units.
Taking the average of Eq. (\ref{W_aux}) with  distribution (\ref{dist}) gives the mean work 
\begin{eqnarray}
    \langle W\rangle =\frac{k_1-k_0}{k_0}\,\frac{T}{2}.
    \label{W_sudden}
\end{eqnarray}
This result also follows from  Eqs. (\ref{work_general}) and (\ref{k_quench}).
According to Eqs. (\ref{W_aux})-(\ref{W_sudden}), 
the normalized  work
$x=W/\langle W\rangle$
is a square of the standard normal random variable $z=\sqrt{k_0/T} q$, $x=z^2$. Therefore $x$ is distributed  with the chi-squared $\chi_1^2$ distribution~\cite{Simon}  
\begin{eqnarray}
f(x)=\frac{1}{\sqrt{2\pi}}  
\frac{e^{-x/2}}{\sqrt{x}}\, \theta(x),
\end{eqnarray}
and the distribution for $W=x\,\langle W\rangle$ has the form 
\begin{eqnarray}
    f(W)=\frac{1}{\sqrt{2\pi\, \langle W\rangle\, W}}\,
    \exp\left(-\frac{W}{2\langle W\rangle}\right)\,\theta\left(\frac{W}{\langle W\rangle}\right).
\label{w_dist_quench}
\end{eqnarray}
This result corroborates Eq. (\ref{w_dist_asymp}), but in contrast to the latter it holds for any $W$.
Thus for the instantaneous quench the work distribution is  non-Gaussian for any values of $W$. The distribution is exactly zero when $W$ has a sign opposite to that of $\langle W\rangle$.

For the instantaneous quench  the validity  of the Jarzynski equality~\cite{Jar} 
\begin{eqnarray}
\langle e^{-W/T}\rangle=e^{-\Delta F/T} 
\label{JE0}
\end{eqnarray}
can be directly verified.
For the oscillator with stiffness $k$ in thermal equilibrium the free energy is 
$F(k)=-T\,\ln(2\pi T\sqrt{m/k})$. Then 
$\Delta F=F(k_1)-F(k_0)=-T\,\ln(\sqrt{k_0/k_1})$,
and the right-hand side of Eq. (\ref{JE0}) equals 
 $\sqrt{k_0/k_1}$. The average in the left-hand side
 can be taken either over $q,v$ with distribution (\ref{dist}), or over $W$ with distribution (\ref{w_dist_quench}):
\begin{eqnarray}
    \langle e^{-W/T}\rangle=
    \iint exp\left(-\frac{k_1-k_0}{2T}\,q^2\right)\,
    \rho_0(q,v)\,\,dq\,dv=\int \exp\left(-\frac{W}{T}\right)\,f(W)\, dW=\sqrt{\frac{k_0}{k_1}}.
\end{eqnarray}

The purpose of this  paper is to consider another limiting and idealized case 
when  
the stiffness is perturbed according to the protocol
\begin{eqnarray}
    k(t)=k_0+a\, \delta(t),
    \label{protocol}
\end{eqnarray}
where $\delta(t)$ is the Dirac delta function and the intensity of the perturbation $a$, which may be of any sign,  has units of mass/time.  We shall refer to Eq. (\ref{protocol}) as the delta protocol.
Of course, 
the  delta protocol cannot be practically realized per se
and should be viewed as an asymptotic form of  spike-like perturbations  
with a  high amplitude and a short duration.
We shall see that the properties of the 
the delta protocol in many respects differ
from that for the instantaneous quench.

\section{Work}
In the sections to follow we show that for the delta protocol (\ref{protocol}) 
the microscopic work on the system is 
\begin{eqnarray}
    W(q,v)=\frac{a^2}{2m}\, q^2 -a\,q\,v,
    \label{W_micro}
\end{eqnarray}
where $q=q(0^-)$ and $v=v(0^-)$ are initial coordinate and velocity just before the perturbation is applied.
Similar to the instantaneous quench, see Eq. (\ref{W_aux}), the statistics of $W$ is completely determined by that of the initial state, 
but for the delta protocol it depends on both initial coordinate 
and velocity. 
The average work is governed by 
the mean square displacement $\langle q^2\rangle$ and
the position-velocity correlation 
$\langle q \,v\rangle$ in the initial state.

Let us assume  that for $t<0$ the system is in thermal equilibrium with distribution (\ref{dist}). In that case the initial correlation
$\langle q v\rangle$ is zero, and the average work
is determined by the initial mean square displacement $\langle q^2\rangle$ only,
\begin{eqnarray}
    \langle W\rangle=\iint W(q,v)\,\rho_0(q,v)\,dq\,dv=\frac{a^2}{2m}\langle q^2\rangle=\frac{a^2T}{2mk_0}=
    \frac{\alpha^2}{2}\, T,\qquad \alpha=\frac{a}{m\,\omega_0}.
    \label{work_delta}
\end{eqnarray}
Here $\omega_0=\sqrt{k_0/m}$ is the oscillator's frequency, and $\alpha$ is the characteristic dimensionless parameter of the protocol.
One observes that $W$ and $\langle W\rangle$  may be, in contrast to the instantaneous quench, of different signs.
Using Eq. (\ref{W_micro}), one can readily evaluate  the moment-generating function
for the dimentionless work $w=W/T$,
\begin{eqnarray}
    M_w(t)=\langle e^{tw}\rangle=\iint e^{t\, W(q,v)/T}\,\rho_0(q,v)\,dq\,dv=\frac{1}{\sqrt{1-
\alpha^2(t+t^2)}}.
\label{mgf}
\end{eqnarray}
Then the moments of $W$ can be evaluated as
\begin{eqnarray}
    \mu_n=\langle W^n\rangle=T^n\,\frac{d^n}{dt^n}\,M_w(t)\bigg\rvert_{t=0}.
\end{eqnarray}
For the first few moments one obtains
\begin{eqnarray}
\mu_1&=&\mu=\frac{1}{2}\,\alpha^2\,T,\nonumber\\
\mu_2&=&\left(
\alpha^2+\frac{3}{4}\,\alpha^4
\right)\,T^2=
2\, T\, \mu+3\,\mu^2, \nonumber\\
    \mu_3&=&
    \left(
    \frac{9}{2}\,\alpha^4
+\frac{15}{8}\,\alpha^6\right)\,T^3=    
    18\, T\,\mu^2+15\,\mu^3, \nonumber\\
    \mu_4 &=&\left(
9\,\alpha^4+\frac{45}{2}\,\alpha^6+
\frac{105}{16}\,\alpha^8\right)
T^4= 
    36\, T^2\, \mu^2+180\, T\,\mu^3 
    +105\,\mu^4,
    \label{moments}
\end{eqnarray}
and the variance $\sigma^2=\mu_2-\mu^2$ equals
\begin{eqnarray}
    \sigma^2=\left(\alpha^2+\frac{1}{2}\,\alpha^4\right)\,T^2=
    2T\mu+2\mu^2.
\end{eqnarray}

From the thermodynamics point of view, the delta protocol is a cycle process,
for which the difference of the free energy is zero, $\Delta F=0$,  and the Jarzynski equality (\ref{JE0}) takes the form $\langle e^{-W/T}\rangle=1$. The validity of this result for the delta protocol is immediately obvious from Eq. (\ref{mgf}),
\begin{eqnarray}
    \langle e^{-W/T}\rangle =M_w(-1)=1.
\end{eqnarray}

For the dimensionless work $w=W/T$, the moment-generating function taken at $-t$ is a bilateral (two-sided) Laplace transform $\mathcal B$ of the distribution function (probability density function)  $f(w)$,
\begin{eqnarray}
    M_w(-t)=
\int_{-\infty}^\infty e^{-t w} f(w)\,dw=\mathcal B \{f(w)\}.
    \label{aux1}
\end{eqnarray}
Then from  Eqs. (\ref{mgf}) and (\ref{aux1}), one finds 
\begin{eqnarray}
      \mathcal B\{f(w)\}= \frac{1}{\sqrt{1+
\alpha^2(t-t^2)}}.
\label{BL}
\end{eqnarray}
The inversion of this relation yields (see the Appendix)
\begin{eqnarray}
    f(w)=\frac{1}{\pi\,|\alpha|}\,
    \exp\left(\frac{w}{2}\right)\, K_0\left(
    \frac{\Delta}{2}\,|w|\right),
    \label{df1}
\end{eqnarray}
where $K_0(x)$ is the modified Bessel function, the parameter $\Delta$ is
\begin{eqnarray}
    \Delta=\sqrt{\frac{4}{\alpha^2}+1},
\end{eqnarray}
and $\alpha$ is defined by Eq. (\ref{work_delta}).
The distribution function for the work $W=T\,w$ takes the form
\begin{eqnarray}
    f(W)=\frac{1}{\pi\,|\alpha|\, T}\,
    \exp\left(\frac{W}{2T}\right)\, K_0\left(
    \frac{\Delta}{2T}\,|W|\right),
    \label{df}
\end{eqnarray}
which is shown in Fig. 1.
As for the instantaneous quench, the work distribution diverges as $W\to 0$. This divergence does not lead to any singularities of  measurable quantities. In particular, using $f(W)$ to evaluate 
the moments
$\mu_n=\int f(W) W^n dW$, one recovers Eqs. (\ref{moments}). Also, the validity of the Jarzynski equality, evaluated with $f(W)$, can be directly verified:
\begin{eqnarray}
    \langle e^{-W/T}\rangle=\int e^{-W/T}\, f(W)\, dW=\int e^{-w}\, f(w)\, dw=1.
\end{eqnarray}

\begin{figure}[t]
  \includegraphics[width=10cm]{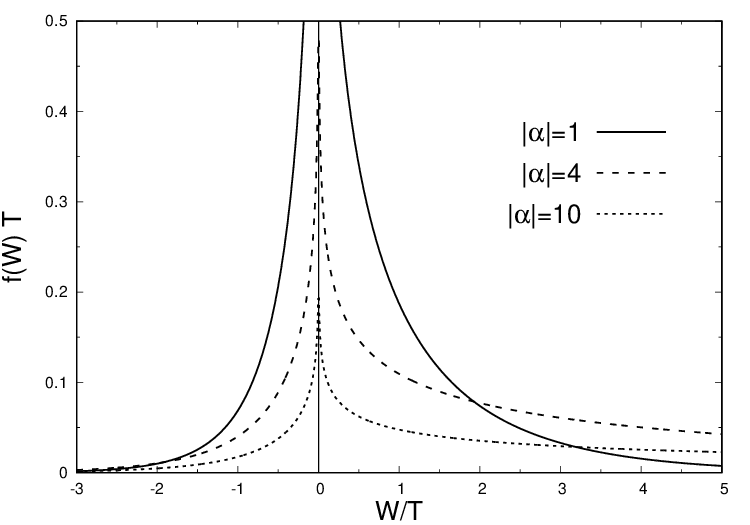}
\caption{The work distribution function (\ref{df}) for several values of $|\alpha|=|a|/(m\,\omega_0)$.
\label{work_dist_fig}}
\end{figure}

Another remarkable, though quite expected, in view of Eq. (\ref{W_micro}), feature of the distribution $f(W)$ is that it is asymmetric.
Taking into account the asymptotic form of the modified Bessel function $K_0(x)\approx\sqrt{\pi/(2 x)}\,e^{-x}$ for large $x$, 
one finds that the asymptotic tails of the distribution for $|W|>2T/\Delta$ have the forms
\begin{eqnarray}
    f(W)\approx
    \frac{1}
    {\alpha\sqrt{\Delta \,\pi\,T\,W}}
    \exp\left(
    -\frac{\Delta-1}{2T}\,W\right), \quad W>0
\end{eqnarray}
and 
\begin{eqnarray}
    f(W)\approx
    \frac{1}
    {\alpha\sqrt{\Delta \,\pi\,T\,|W|}}
    \exp\left(
    \frac{\Delta+1}{2T}\,W\right), \quad W<0.
\end{eqnarray}
The distribution decays faster for negative values of $W$. If in addition $|\alpha|\gg 1$, then $\Delta\to 1$, $\Delta-1\to 2/\alpha^2$, and the above expressions are further simplified:
\begin{eqnarray}
    f(W)\approx
    \frac{1}
    {\sqrt{2\,\pi\,\langle W\rangle \,W}}
    \exp\left(
    -\frac{W}{2\langle W\rangle}\right), \quad W>0,
    \label{asymp3}
\end{eqnarray}
and 
\begin{eqnarray}
    f(W)\approx
    \frac{1}
    {\sqrt{2\,\pi\,\langle W\rangle \,|W|}}
    \exp\left(
    \frac{W}{T}\right), \quad W<0.
\end{eqnarray}
Note that  the asymptotic expression (\ref{asymp3}) coincides with the exact distribution for the instantaneous quench, Eq. (\ref{w_dist_quench}).

Whereas the derivation of the above results  
is quite straightforward, there is a subtlety
related to the fact that for the delta protocol the microscopic velocity is  
discontinuous at $t=0$, see Eq. (\ref{v_discont}) below.
The next section elaborates on that.

\section{Preliminaries}
For the delta protocol (\ref{protocol}) 
Eq. (\ref{work_general}) for the average work takes the form
\begin{eqnarray}
\langle W\rangle=
    \frac{a}{2}\,\int_{-\epsilon}^{\epsilon}  \langle q^2(t)\rangle\,\dot \delta(t)\, dt,
    \label{work_general2}
\end{eqnarray}
with $\epsilon>0$.
Integrating by parts yields
\begin{eqnarray}
\langle W\rangle=-\frac{a}{2}\,\int_{-\epsilon}^{\epsilon}\!\delta(t)\,\frac{d}{dt}\,\langle q^2(t)\rangle\,dt. 
    \label{work_general3}
\end{eqnarray}
As a next step, it might be tempting to apply
the defining property of the delta function 
\begin{eqnarray}
    \int_{-\epsilon}^\epsilon \delta (t) f(t) dt=f(0)
    \label{delta}
\end{eqnarray}
to obtain
\begin{eqnarray}
\langle W\rangle=
    -\frac{a}{2}\,\,\frac{d}{dt}\,\langle q^2(t)\rangle\bigg\rvert_{t=0}=-a\,\langle q(0)\, v(0)\rangle. 
    \label{error}
\end{eqnarray}
Suppose the system is in thermal equilibrium at $t<0$, so that 
the position-velocity correlation $\langle q(t)\,v(t)\rangle$ is zero at $t=0^-$.
Assuming that the correlation  is continuous at $t=0$, one would find  
from  Eq. (\ref{error}) that 
the mean work is zero, $\langle W\rangle=0$.

That conclusion is incorrect because, as  we shall see below, the function 
\begin{eqnarray}
    f(t)=\frac{d}{dt}\,\langle q^2(t)\rangle=2\,\langle q(t) \,v(t)\rangle
    \label{f}
\end{eqnarray}
is discontinuous at $t=0$. Since $f(0)$ is not well-defined, 
the property (\ref{delta}) is ambiguous
and
the transition 
from Eq. (\ref{work_general3}) to Eq. (\ref{error}) is in general erroneous.
The standard way to extend the property (\ref{delta}) for the
case of discontinuous $f(t)$ is to replace $f(0)$ by the average of $f(t)$ at the 
discontinuity
\begin{eqnarray}
    \int_{-\epsilon}^\epsilon \delta (t) f(t) dt=\frac{1}{2}\left[f(0^-)+f(0^+)\right].
    \label{delta2}
\end{eqnarray}
In our case $f(0^-)=0$ (if the system is in equilibrium at $t<0$) and $f(0^+)$  will be shown 
below, see Eq. (\ref{f_right}), to be
\begin{eqnarray}
    f(0^+)=-\frac{2 a T}{m k_0}.
\end{eqnarray}
Then from Eqs. (\ref{work_general3}) and (\ref{delta2}) one gets
\begin{eqnarray}
    \langle W\rangle=-\frac{a}{4}\,f(0^+)=\frac{a^2}{m\,k_0}\, \frac{T}{2},
\end{eqnarray}
which is the result (\ref{work_delta}).

The ansatz (\ref{delta2}) is known to be inconsistent in certain cases 
and ought to be used with care~\cite{GW}.
Below we consider the problem with two methods.  The first method is based on using  
ansatz (\ref{delta2}). 
The second method does not rely on  ansatz (\ref{delta2}), but involves a specific representation of $\delta(t)$  as a limit of a rectangular 
impulse. 
We find that the two methods lead to the same results.

\section{Method I}
According to Eq. (\ref{W_general0}), for the delta protocol the microscopic work is 
\begin{eqnarray}
    W=\frac{a}{2}\,\int_{-\epsilon}^\epsilon \! q^2(t)\,\dot \delta (t) \,dt=-a\int_{-\epsilon}^\epsilon \! q(t)\, v(t)\, \delta (t)\,dt. 
    \label{work_method2}
\end{eqnarray}
Anticipating $q(t)$ and $v(t)$ to be, respectively, continuous and discontinuous at $t=0$, and applying ansatz (\ref{delta2}) 
one gets
\begin{eqnarray}
    W=-\frac{a}{2}\, q(0)\,[v(0^+)+v(0^-)].
    \label{aux2}
\end{eqnarray}
Here $q=q(0)$ and $v=v(0^-)$  are drawn from the initial distribution, which may be, but not necessarily,  the equilibrium distribution (\ref{dist}). 
In order to find $v(0^+)$ we need to solve equations of stochastic dynamics, find $v(t)$ for $t>0$, and
take the limit $t\to 0^+$.

To this end, we assume that for $t>0$ 
the system is described by the generalized Langevin equation~\cite{Zwanzig,Weiss}
\begin{eqnarray}
\!\!
m\,\ddot q(t)\!=\!-k(t)\,q(t)\!-\!m \int_0^t \!K(t-\tau)\,\dot q(\tau)\, d\tau+\xi(t).
\label{gle}
\end{eqnarray}
Here $\xi(t)$
is the stationary fluctuating
force (``noise'') which is zero-centered, $\langle \xi(t)\rangle=0$,
and related to the dissipation kernel $K(t)$
via the fluctuation-dissipation relation 
\begin{eqnarray}
\langle \xi(t)\,\xi(t')\rangle=m\,T\,K(|t-t'|).
\label{fdr}
\end{eqnarray}
We do not assume that $\xi(t)$ is Gaussian.
The anticipated results (\ref{W_micro}) and (\ref{work_delta}) do not depend on the form of the dissipation kernel $K(t)$. Therefore, instead of the generalized Langevin equation one can use its more simple and  familiar Markovian form with $K(t)=\gamma\, \delta(t)$.  Yet we prefer to work with the Langevin equation of the form (\ref{gle}) which is more general and,  
in contrast to the Markovian  
counterpart, does not imply  coarse-graining of time. We shall show in Conclusion that the 
overdamped approximation (when the acceleration term is put to zero) is inadequate to describe the delta protocol.

The Langevin equation (\ref{gle}) for 
protocol (\ref{protocol}) takes the form
\begin{eqnarray}
\ddot q(t)=-\left[\omega_0^2+\frac{a}{m}\,\delta(t)\right]\,q(t)
-\int_0^t K(t-\tau)\,\dot q(\tau)\, d\tau+\frac{1}{m}\xi(t),
\label{gle2}
\end{eqnarray}
where $\omega_0^2=k_0/m$ is the oscillator's natural frequency.
The equation is linear and can be readily solved using Laplace transforms
\begin{eqnarray}
    \mathcal L \{f(t)]=\tilde f(s)=\int_{0^-}^\infty e^{-st} f(t)\,dt.
\end{eqnarray}
Assuming that $q(t)$  (but not $v(t)=\dot q(t)$) is continuous at $t=0$, 
one finds
\begin{eqnarray}
q(t)=S_0(t)\, q(0)+G_0(t)
\left[v(0^-)\!-\!\frac{a}{m} q(0)
\right]+\frac{1}{m}\{G_0*\xi\}.
\label{sol}
\end{eqnarray}
Here the asterisk in the last term stands for the convolution
$\{G_0*\xi\}=\int_0^t G_0(t-\tau)\, \xi(\tau)\, d\tau$, 
and the relaxation functions $G_0(t)$ and $S_0(t)$ are defined in the Laplace domain as follows, 
\begin{eqnarray}
\tilde G_0(s)=\frac{1}{s^2+s\tilde K +\omega_0^2},\quad
\tilde S_0(s)=\frac{s+\tilde K}{s^2+s\tilde K+\omega_0^2}=\frac{1}{s}\,\left[1-\omega_0^2\,\tilde G_0(s)\right].
\label{GS}
\end{eqnarray}
It is also convenient to use another relaxation function $R_0(t)$, which is connected to $G_0(t)$ in the Laplace domain as
\begin{eqnarray}
    \tilde R_0(s)=s\,\tilde G_0(s).
    \label{R}
\end{eqnarray}
One can show that the relaxation functions coincide with the normalized 
correlations for the oscillator with $k=k_0$ in thermal equilibrium~\cite{Wang,DV,Goychuk,Plyukhin}
\begin{eqnarray}
    \langle q(t) q(t')\rangle_{eq}=\frac{T}{k_0}\, S_0(t-t'),\quad
\langle v(t) v(t')\rangle_{eq}=\frac{T}{m}\, R_0(t-t'),\quad
\langle q(t) v(t')\rangle_{eq}=\frac{T}{m}\, G_0(t-t').
\end{eqnarray}
From these relations it is  clear that the relaxation functions are continuous at $t=0$, and
their initial values are
\begin{eqnarray}
S_0(0)=R_0(0)=1, \quad G_0(0)=0, \quad
    \dot S_0(0)=\dot R_0(0)=0, \quad \dot G_0(0)=1.
    \label{relaxation_initial}
\end{eqnarray}
These initial values can also be obtained directly using the definition of the relaxation
functions in the Laplace domain, Eqs. (\ref{GS}) and (\ref{R}). 

Using the above relations,  one finds that in the time domain the relaxation functions are connected as follows
\begin{eqnarray}
    \dot G_0(t)=R_0(t), \quad \dot S_0(t)=-\omega_0^2 G_0(t).
    \label{derivatives}
\end{eqnarray}
Then differentiating Eq. (\ref{sol}) yields
\begin{eqnarray}
   v(t)= -q(0)\,\left[\omega_0^2G_0(t)+\frac{a}{m}R_0(t)\right]
   +v(0^-)\,R_0(t)
   +\frac{1}{m}\,\{R_0*\xi\}.
   \label{v}
\end{eqnarray}
Taking in Eqs. (\ref{sol}) and (\ref{v}) the limit $t\to 0^+$ and taking into account Eq. (\ref{relaxation_initial}) one finds that at $t=0$ the coordinate $q(t)$ 
is continuous (in accordance with
our initial assumption), while  the velocity $v(t)$
has a jump discontinuity (unless $q(0)=0$):
\begin{eqnarray}
    v(0^+)=v(0^-)-q(0)\,\frac{a}{m}. \label{v_discont}
\end{eqnarray}
This relation can also be obtained directly by integrating the Langevin equation (\ref{gle2}) over $t$ from $0^-$ to $0^+$.
Substituting Eq. (\ref{v_discont}) into Eq. (\ref{aux2}) yields for the microscopic work the result (\ref{W_micro}),
\begin{eqnarray}
    W=\frac{a^2}{2m}\,q^2 -a\, q\, v,
\end{eqnarray}
where $q=q(0)$ and velocity $v=v(0^-)$ 
are the coordinate and velocity immediately before the perturbation.

\section{Energy}
Suppose the system at $t<0$ is in thermal equilibrium with the distribution $\rho_0$, Eq. (\ref{dist}).  
For the instantaneous quench $k_0\to k_1$ at $t=0$
the average kinetic energy $E_k=m\langle v^2(t)\rangle/2$ can be shown to be continuous and the average potential energy $E_p=k(t)\langle q^2(t)\rangle/2$ has the instantaneous gain:
\begin{eqnarray}
    E_k(0^+)=E_k(0^-),\quad E_p(0^+)=E_p(0^-)+\langle W\rangle.
\end{eqnarray}
Here $E_k(0^-)=E_p(0^-)=T/2$ and $\langle W\rangle$ is the mean work on the system given by Eq. (\ref{W_sudden}).  In other words, immediately after the quench the work is adopted by the system  entirely in the form of potential energy. 

One can show  that 
for the delta protocol the situation is, in a sense, opposite: At $t=0$ the average potential energy is continuous, whereas the kinetic energy receives instantaneously  a finite gain,
\begin{eqnarray}
    E_p(0^+)=E_p(0^-),\quad E_k(0^+)=E_k(0^-)+\langle W\rangle.\label{Epk}
\end{eqnarray}
Here again $E_k(0^-)=E_p(0^-)=T/2$ and $\langle W\rangle$ is the mean work on the system given by Eq. (\ref{work_delta}). Thus, immediately after the delta perturbation  the potential energy is unaffected, and the work on the system is utilized as  the system's kinetic energy.

The average potential energy is determined by 
$\langle q^2(t)\rangle$.
Squaring Eq. (\ref{sol}) and  taking the average with the equilibrium distribution $\rho_0$ yields
\begin{eqnarray}
\langle q^2(t)\rangle=\frac{T}{k_0} \left\{
S_0(t)-\frac{a}{m} G_0(t)
\right\}^2+\frac{T}{m}\,G_0^2(t)+\frac{1}{m^2}\,\langle (G_0*\xi)^2\rangle.
\label{aux5}
\end{eqnarray}
The average square of the convolution in this expression can be,   
using the fluctuation-dissipation relation (\ref{fdr}), 
worked out to the form
\begin{eqnarray}
\langle (G_0*\xi)^2\rangle=\frac{m\,T}{\omega_0^2}\,[1-S_0^2(t)]-m\,T\,G_0^2(t).
\label{conv}
\end{eqnarray}
Then  we get
\begin{eqnarray}
\langle q^2(t)\rangle=\frac{T}{k_0}\left\{
1-\frac{2a}{m}\,S_0(t)\,G_0(t)+\left(\frac{a}{m}\right)^2G_0^2(t)
\right\}.
\label{q2}
\end{eqnarray}
This expression holds for $t>0$
Taking the limit $t\to 0^+$ and taking into account that $G_0(0)=0$, see Eq. (\ref{relaxation_initial}), one finds
\begin{eqnarray}
    \langle q^2(0^+)\rangle =\langle q^2(0^-)\rangle=T/k_0.
\end{eqnarray}
This leads to the continuity of the potential energy.

Although the mean square displacement is continuous, its derivative,  
which is also the two times 
position-velocity correlation,
\begin{eqnarray}
    f(t)=\frac{d}{dt}\,\langle q^2(t)\rangle =2 \langle q(t) v(t)\rangle,
\end{eqnarray}
is discontinuous at $t=0$.  Indeed, from Eq. (\ref{q2}) and (\ref{relaxation_initial}) one finds
\begin{eqnarray}
    f(0^+)=\lim_{t\to 0^+} \frac{d}{dt}\,\langle q^2(t)\rangle=-\frac{2 a T}{m k_0}.
\label{f_right}
\end{eqnarray}
On the other hand, for $t<0$ the system is in equilibrium, so that $f(0^-)=0$, and
\begin{eqnarray}
    f(0^+)=f(0^-)-\frac{2 aT}{mk_0}.
\end{eqnarray}
We referred to the discontinuity of $f(t)$ is Sec. III, evaluating the mean work using Eq. (\ref{work_general3}).

Consider now the average kinetic energy. 
Squaring and averaging of Eq. (\ref{v}) yields
\begin{eqnarray}
    \langle v^2(t)\rangle&=&\frac{T}{k_0}\,\left\{\omega_0^2G_0(t)+\frac{a}{m} R_0(t)\right\}^2+\frac{T}{m}\,R_0^2(t)
    +\frac{1}{m^2}\,\langle (\xi*R_0)^2\rangle.
\end{eqnarray}
Similar to Eq. (\ref{conv}),  
the last term can be worked out to the form
\begin{eqnarray}
\langle (R_0*\xi)^2\rangle=m\,T\,[1-R_0^2(t)-\omega_0^2\,G_0^2(t)],
\label{conv2}
\end{eqnarray}
then 
\begin{eqnarray}
\!\!\!
\langle v^2(t)\rangle=\frac{T}{m}\left\{
1\!+\!\frac{2a}{m}\,G_0(t)\,R_0(t)\!+\!\left(\frac{a}{m\,\omega_0}\right)^2\!\!R_0^2(t)
\right\}.
\label{v2}
\end{eqnarray}
Taking the limit $t\to 0^+$ we find that the velocity's second moment is discontinuous, 
\begin{eqnarray}
    \langle v^2(0^+)\rangle=\langle v^2(0^-)\rangle+\left(\frac{a}{m\omega_0}\right)^2\frac{T}{m}, 
    \label{v2_drop}
\end{eqnarray}
where $\langle v^2(0^-)\rangle =T/m$.
Written for the average kinetic energy
$E_k(t)=m \langle v^2(t)\rangle /2$, this discontinuity relation takes the form
Eq. (\ref{Epk}) and leads to an 
observation  that  the mean work $\langle W\rangle$ performed on the system 
is collected by the system entirely in the form of the kinetic energy.

\section{Method II}
One may worry about the validity of the method I exploited in Sec. IV,  because it takes  for granted the ad hoc ansatz
(\ref{delta2}) for the discontinuous function $f(t)=\frac{d}{dt}\langle q^2(t)\rangle=2\langle q(t)\,v(t)\rangle$. Also, solving the Langevin equation (\ref{gle}) using the Laplace transformation ($\mathcal L$), we tacitly used the relation  $\mathcal L\{\delta (t)\, q(t)\}=q(0)$.  This  implies  the continuity of  $q(t)$ at $t=0$, which is not obvious for the delta protocol.

In this section we apply another method which does not rely 
on the ansatz (\ref{delta2}) and does not assume the continuity of $q(t)$,  but instead 
adopts a specific representation of the delta function.
Namely, we assume that the stiffness $k(t)$ is varied according to the piecewise-constant protocol
\begin{eqnarray}
k(t) =
   \begin{cases}
     k_0, & t<0, \\
     k_1=k_0+a/\epsilon, & 0\le t<\epsilon,\\
     k_0, & t\ge\epsilon,
   \end{cases}
   \label{protocol2}
\end{eqnarray}
where $\epsilon, a>0$. 
The perturbation has the form of a rectangular impulse of duration $\epsilon$ and amplitude $a/\epsilon$.
In the limit $\epsilon\to 0$, which will be applied later,  protocol (\ref{protocol2})
converges to the delta protocol.

The protocol (\ref{protocol2}) can be viewed as a sequence of two instantaneous quenches: 
$k_0\to k_1$ at $t=0^-$ and $k_1\to k_0$ at $t=\epsilon$. 
Accordingly, 
the microscopic work has two contributions, 
\begin{eqnarray}
W =W_{01} 
+W_{10}, 
\label{work}
\end{eqnarray}
where, similar to Eq.(\ref{W_aux}), 
\begin{eqnarray}
W_{01}=
\frac{k_1-k_0}{2}\,q^2(0^-)
=\frac{a}{2\epsilon}\,q^2(0^-), \quad W_{10}=
\frac{k_0-k_1}{2}\,q^2(\epsilon)=
-\frac{a}{2\epsilon}\,q^2(\epsilon).
\label{aux12}
\end{eqnarray}
In order to find  $q^2(\epsilon)$, one
needs to solve the Langevin equation (\ref{gle}) for $t>0$  with $k(t)=k_1$
\begin{eqnarray}
\ddot q(t)=-\omega_1^2\,q(t)-\int_0^t K(t-\tau)\,\dot q(\tau)\, d\tau+\frac{1}{m}\xi(t),
\label{gle3}
\end{eqnarray}
where $\omega_1^2=k_1/m$. Solving the equation  with Laplace transforms, one finds
\begin{eqnarray}
    q(t)=q(0^-)\, S_1(t)+v(0^-)\, G_1(t)+L_1(t).
    \label{sol2}
\end{eqnarray}
Here $L_1(t)$ denotes the convolution 
\begin{eqnarray}
    L_1(t)=\frac{1}{m}\,\{G_1*\xi\}=\frac{1}{m}\,\int_0^t G_1(t-t') \xi(t')\,dt',
\end{eqnarray} 
and 
the relaxation functions $S_1$ and $G_1$ are defined as in the previous section but now they corresponds to the stiffness $k_1$ and frequency $\omega_1=\sqrt{k_1/m}$.
Then for the microscopic work  
$W=W_{01}+W_{10}$ we obtain 
\begin{eqnarray}
    W&=&\frac{a}{2\epsilon}\,[q^2- q^2(\epsilon)]\nonumber\\
    &=&\frac{a}{2\epsilon}\left\{
    [1-S_1^2(\epsilon)]\,q^2-G_1^2(\epsilon)\,v^2-L_1^2(\epsilon)
    \right\}
    -\frac{a}{\epsilon}\left\{
    S_1(\epsilon)G_1(\epsilon)\,q\,v-[S_1(\epsilon)\,q+G_1(\epsilon)\,v]\,
    L_1(\epsilon)
    \right\}.
    \label{WW1}
\end{eqnarray}
Here and below for brevity we use the notations $q=q(0^-)$, $v=v(0^-)$.

We are looking for the asymptotic form of the expression  (\ref{WW1}) in the limit $\epsilon\to 0$,
when the protocol (\ref{protocol2}) converges to the delta protocol.
According to Eqs. (\ref{relaxation_initial}) and (\ref{derivatives}) (in which one needs to replace the subscript $0$ by $1$), the initial values of the relevant functions are 
\begin{eqnarray}
    && G_1(0)=0, \quad \dot G_1(0)=1, \quad \ddot G_1(0)=0\nonumber\\
    && S_1(0)=1, \quad \dot S_1(0)=0,\quad \ddot S_1(0)=-\omega_1^2\approx -\frac{a}{m\epsilon}\nonumber\\
    && L_1(0)=0,\quad \dot L_1(0)=0,\quad \ddot L_1(0)=\frac{\xi(0)}{m}.
    \label{relax_zero}
\end{eqnarray}
Then at small $t=\epsilon$ 
the relaxation functions have the following  asymptotic forms
\begin{eqnarray}
G_1(\epsilon)=\epsilon+\mathcal O(\epsilon^2),\quad
    S_1(\epsilon)=1-\frac{a}{2m}\,\epsilon+\mathcal O(\epsilon^2),\quad
    S_1^2(\epsilon)=1-\frac{a}{m}\epsilon+\mathcal O(\epsilon^2),\quad
    L_1(\epsilon)=\mathcal O(\epsilon^2).
\end{eqnarray}
Substituting into Eq. (\ref{WW1})  yields to the leading (zero) order in $\epsilon$, i.e. in the limit $\epsilon\to 0$, the expression 
\begin{eqnarray}
    W=\frac{a^2}{2m}\, q^2-a\,  q\, v.
\label{WW3}
\end{eqnarray}
This coincides with the result we obtained in Sec. IV with method I.

\section{Response}
The system's response  to the delta perturbation, i.e. the dynamics for $t>0$,  
is more convenient to consider with method I by solving 
Langevin equation (\ref{gle2}). Let us focus on 
the mean square displacement $\langle q^2(t)\rangle$, which determines the response of the system's potential energy.   

Let us assume that at $t<0$ the system is in thermal equilibrium with the distribution $\rho_0$, Eq. (\ref{dist}).
In Section V  we obtained for $\langle q^2(t)\rangle$ the result Eq. (\ref{q2}).
Let us re-write it as
\begin{eqnarray}
\langle q^2(t)\rangle=\varphi(t) \,\langle q^2(0^-)\rangle,
\label{response_def}
\end{eqnarray}
where $\langle q^2(0^-)\rangle=T/k_0$ is the equilibrium mean-square displacement (before the perturbation is applied) and the dimensionless response function $\varphi(t)$
has the form 
\begin{eqnarray}
\varphi(t)=
1-2\alpha\,S_0(t)\,\omega_0 G_0(t)+
\alpha^2\, [\omega_0 G_0(t)]^2.
\label{response}
\end{eqnarray}
Since $G_0(0)=0$, the initial value of $\varphi(t)$ is $1$ for any memory kernel $K(t)$ in the Langevin equation.    The behavior of $\varphi(t)$ for $t>0$
depends on a particular form of the dissipation kernel $K(t)$. Below we consider two examples for which the relaxation functions $G_0(t)$ and $S_0(t)$ are available in closed form.

\begin{figure}[t]
\includegraphics[width=18cm]{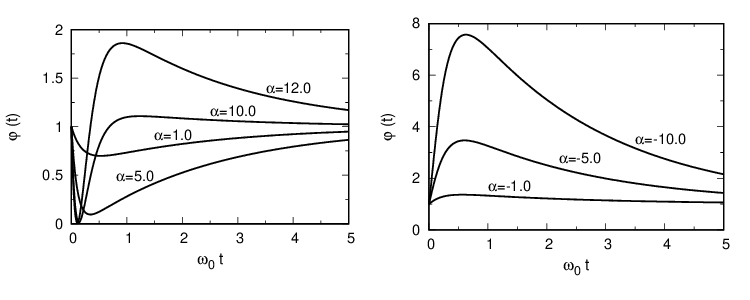}
\caption{The response function $\varphi(t)$, defined by Eq. (\ref{response_def}), for the 
Markovian overdamped oscillator, described by the generalized Langevin equation with the kernel $K(t)=\gamma\,\delta(t)$, for $\beta=\gamma/\omega_0=5.0$ and 
several positive (left) and negative (right) values of the perturbation
intensity $\alpha=a/(m\omega_0)$. }
\label{response1_fig}
\end{figure}

As the first example consider the kernel
\begin{eqnarray}
    K(t)=\gamma\,\delta(t), \qquad  \gamma^2-4 \omega_0^2>0,
    \label{K_delta}
\end{eqnarray}
which corresponds to the Markovian Langevin equation for the  overdamped  oscillator. According to Eq. (\ref{GS}), the relaxation functions have the forms
\begin{eqnarray}
    \omega_0\,G_0(t)=\frac{1}{\sqrt{\beta^2-4}}\left\{
    e^{-s_1 (\omega_0t)}-e^{-s_2 (\omega_0t)}
    \right\}, \quad
    S_0(t)=\frac{1}{\sqrt{\beta^2-4}}\,
    \left\{
    \frac{1}{s_1}\,e^{-s_1 (\omega_0t)}-
     \frac{1}{s_2}\,e^{-s_2(\omega_0t)}\right\},
     \label{green1}
\end{eqnarray}
where dimensionless parameters $\beta,s_1,s_2$ are
\begin{eqnarray}
  \beta=\frac{\gamma}{\omega_0}>2, \quad
  s_1=\frac{1}{2}\,(\beta-\sqrt{\beta^2-4}),\quad
  s_2=\frac{1}{2}\,(\beta+\sqrt{\beta^2-4}).
\end{eqnarray}
The corresponding response function $\varphi(t)$, Eq. (\ref{response}), for $\beta=5.0$
is shown in Fig.~\ref{response1_fig} .
For positive $\alpha$, the behavior of $\varphi(t)$ is remarkably different for smaller and larger values of $\alpha$. For  $\alpha<\alpha_0\approx 9.6$, $\varphi(t)$ first decreases, reaches a minimum, and then monotonically increases back to $1$. For $\alpha>\alpha_0$, after reaching a minimum, $\varphi(t)$ develops a maximum, and then monotonically decreases back to $1$. 
For $\alpha<0$, $\varphi(t)$ behaves in a similar way for all values of $\alpha$:
It first increases,
reaches a maximum, and then monotonically decreases back to the initial value $1$.

Our second example is the kernel
\begin{eqnarray}
    K(t)=\frac{\omega_*}{2}\,\frac{J_1(\omega_* t)}{t}=\frac{\omega_*^2}{4}[J_0(\omega_* t)+J_2(\omega_* t)],
\label{K}
\end{eqnarray}
where $J_n(t)$ are  the Bessel functions of the first kind. 
The kernel has the absolute maximum at $t=0$ and for $t>0$
it oscillates with an amplitude decaying with time as $t^{-3/2}$.
The generalized Langevin equation with kernel (\ref{K}) 
describes  
a particle attached to a semi-infinite harmonic chain, which plays the role of the thermal bath~\cite{Weiss}, for the case when the mass of the particle and the masses of chain particles 
are the same.
The frequency $\omega_*$ has the meaning of the maximal frequency of the normal modes of the bath.

\begin{figure}[t]
  \includegraphics[width=18cm]{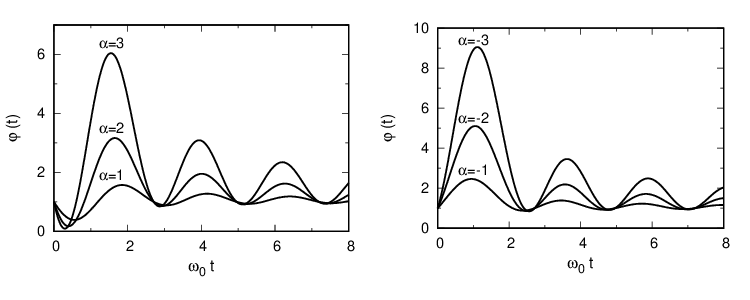}
\caption{The response function $\varphi(t)$ given by  (\ref{response_explicit}), 
for several positive (left) and negative (right) values of the perturbation
intensity $\alpha=a/(m\omega_0)$.}
\label{response2_fig}
\end{figure}

The oscillator described by the generalized Langevin equation with kernels like Eq. (\ref{K})
was studied in Ref.~\cite{Plyukhin}. Here 
we  consider only the special case when
the oscillator frequency $\omega_0$ and the maximal bath frequency $\omega_*$ are related as 
\begin{eqnarray}
    \omega_0=\omega_*/\sqrt{2}.
\label{omega}
\end{eqnarray}
In that case the relaxation functions can be presented in closed form as follow
\begin{eqnarray}
    G_0(t)=\frac{2}{\omega_*}\,J_1(\omega_* t)=\frac{\sqrt{2}}{\omega_0}\,J_1(\sqrt{2}\,\omega_0\, t),
    \quad
    S_0(t)=J_0(\omega_* t)=J_0(\sqrt{2}\,\omega_0\, t).
    \label{green2}
\end{eqnarray}
Then the response function (\ref{response}) takes the form
\begin{eqnarray}
    \varphi(t)=1-2\sqrt{2}\,\alpha\,J_0(\sqrt{2}\,\omega_0 t)\,J_1(\sqrt{2}\,\omega_0 t)
    +2\,\alpha^2 J_1^2(\sqrt{2}\,\omega_0 t).
    \label{response_explicit}
\end{eqnarray}
Fig.~\ref{response2_fig} shows that the response function has the form of 
oscillations with the amplitude decreasing with time. For small values of $\alpha$ the oscillation amplitude is small
 too, so  that the perturbed mean square displacement
$\langle q^2(t)\rangle$  remains close to the equilibrium value. For larger values of $\alpha$ 
the minimal value of $\varphi(t)$  is getting closer to $0$, whereas the maximum value increases.  
Suppose the delta perturbation is followed by 
an instantaneous quench $k_0\to k_1$. By choosing the moment of the quench $t_*$ to coincide
with the time of absolute maximum (minimum)
of the response function $\varphi(t)$, the mean work for the quench 
$\frac{1}{2}(k_1-k_0)\langle q^2(t_*)\rangle$ can be strongly increased (decreased), compared to the case when the quench disturbs the system in equilibrium.

\section{Conclusion}

A special property of the instantaneous quench and delta protocols is that the microscopic work
in both cases does not depend on parameters describing stochastic dynamics, but is completely 
determined by equilibrium statistics of the initial state. Intuitively, that might be expected on the ground that both  protocols have effectively zero duration. However, in contrast to the instantaneous quench, in order to evaluate  the work for the delta protocol one does need to take stochastic dynamics  into account, though only on the asymptotically short time scale.
On that scale the properties of the relevant relaxation (correlation) functions are generic and do not depend on the specific form of the 
dissipation kernel $K(t)$ and statistics of noise in the generalized Langevin equation. 

From that perspective it is clear that 
the approximation of overdamped Langevin dynamics, which implies a coarse-graining of time, is inadequate to describe the delta protocol. The solution of the overdamped Langevin equation for the unperturbed oscillator 
\begin{eqnarray}
    \gamma\,\dot q(t)=-\omega^2\,q(t)+\frac{1}{m}\,\xi(t)
\end{eqnarray}
is expressed in term of the exponential 
relaxation function $S(t)=\exp(-\omega^2 t/\gamma)$. The derivative of the latter at $t=0$ is finite, whereas
beyond the overdamped approximation the derivative of the corresponding relaxation function is strictly zero,
$\dot S(0)=0$, see Eq. (\ref{relax_zero}) and also Eqs. (\ref{green1}) and (\ref{green2}) for specific examples.
The difference is essential.
Repeating the arguments of Section VI one finds that for the overdamped case the work diverges as the width $\epsilon$ of the perturbation impulse goes to zero, which is, of course, an unphysical result.

It is but natural to ask
to what extent the idealized delta protocol can mimic realistic spike-like perturbations 
with finite duration and amplitudes. 
Consider, for instance,  a Gaussian protocol
\begin{eqnarray}
    k(t)=k_0+
    \frac{a}{\epsilon\,\sqrt{\pi}}\,e^{-(t/\epsilon)^2},
    \label{gauss}
\end{eqnarray}
which converges to the delta protocol when $\epsilon\to 0^+$.
In that case solving the Langevin equation (\ref{gle}) analytically 
appears to be not feasible. Instead, we can take into account that the Langevin equation 
with the kernel (\ref{K}) describes the terminal atom of a semi-infinite harmonic chain and 
to study the response of this system with numerical simulation. In that way 
we found that Eq. (\ref{response_explicit}) describes the response reasonably well 
already for $\omega_*\epsilon=0.1$ (underestimating  local maxima of the response function by less than ten per cent). 
For 
$\omega_*\epsilon<0.01$ we found that the response function $\varphi(t)$ and the mean work $\langle W\rangle$ for the delta and Gaussian protocols are practically indistinguishable.   

Consider the specific setting with
$\omega_0=\omega_*/\sqrt{2}$ and $\alpha=a/(m\omega_0)=\sqrt{2}$. For the delta protocol the first two moment of the dimensionless work
$w=W/T$ are $\langle w\rangle=1$ and $\langle w^2\rangle=5$, see Eq. (\ref{moments}).
For the Gaussian protocol (\ref{gauss}) the simulation (of about $10^4$ runs) gives
the values
$\langle w\rangle=\{0.99,\, 0.96, \,0.91 \}$ and
$\langle w^2\rangle=\{4.94, \,4.68,\, 4.3 \}$
for $\omega_*\epsilon=\{0.01, \,0.05,\,0.1\}$, 
respectively.

In this paper we assumed that dynamics of the system for $t>0$ is governed by the generalized Langevin equation (\ref{gle}). Its derivations usually  assume that at $t=0^-$ the bath has equilibrated in the presence of the system. An alternative is  that the bath is equilibrated separately, i.e. in the absence of the system. In that case  the Langevin equation still can be derived under certain model assumptions~\cite{Zwanzig,Weiss},  but now it has somewhat different form
\begin{eqnarray}
m\,\ddot q(t)\!=\!-k(t)\,q(t)\!-\!m \int_0^t \!K(t-\tau)\,\dot q(\tau)\, d\tau+\xi(t)- m\,K(t)\,q(0),
\label{gle4}
\end{eqnarray}
which involves  
the additional term in the right-hand side
depending on $q(0)$ (often referred to as the initial slip).
Using the same methods as above, one can show that 
the Langevin equation (\ref{gle4}) leads again to the result (\ref{W_micro}) for the work.

\renewcommand{\theequation}{A\arabic{equation}}
  \setcounter{equation}{0}  

  \section*{APPENDIX}  

In this Appendix we obtain 
the probability distribution function $f(w)$ 
for the dimensionless work $w=W/T$.
The relation between $f(w)$ and 
the moment-generating function $M_w(t)$  
has the form
 \begin{eqnarray}
     \mathcal B\{f(w)\}=\int_{-\infty}^\infty e^{-t w} f(w)\,dw=M_w(-t),
     \label{w1}
 \end{eqnarray}
 where $\mathcal B$ denotes the bilateral Laplace transform.
 Evaluating the right hand-side as
 \begin{eqnarray}
    M_w(-t)=\langle e^{-tw}\rangle=\iint e^{-t\, W(q,v)/T}\,\rho_0(q,v)\,dq\,dv,
\label{w2}
\end{eqnarray}
with $\rho_0$ and $W$ given by Eqs. (\ref{dist}) and (\ref{W_micro}), one finds
\begin{eqnarray}
    M_w(-t)=\frac{1}{\sqrt{1+\alpha^2(t-t^2)}},\quad \alpha=\frac{a}{m\omega_0},
\label{w3}
\end{eqnarray}
provided the argument of the square-root is positive $1+\alpha^2(t-t^2)>0$. The latter condition can also be written as
\begin{eqnarray}
t_1<t<t_2,
\label{w4}
\end{eqnarray}
where
\begin{eqnarray}
\quad t_1=\frac{1-\sqrt{4/\alpha^2+1}}{2}<0, 
    \quad 
     t_2=\frac{1+\sqrt{4/\alpha^2+1}}{2}>0.
     \label{w5}
     \end{eqnarray}
From the above relations we find that the bilateral Laplace transform of $f(w)$ is
 \begin{eqnarray}
         \mathcal B\{f(w)\}= \frac{1}{\sqrt{1+
\alpha^2(t-t^2)}}=\frac{1}{|\alpha|\,\sqrt{(t-t_1)(t_2-t)}},
     \label{w6}
 \end{eqnarray}
 and the region of convergence of the transform 
 has the form (\ref{w4}).
The inverse transformation involves analytical continuation of the above expression to the complex plane and the evaluation of the Bromwich integral 
\begin{eqnarray}
    f(w)=\frac{1}{2\pi i}\,\int\limits_{\gamma-i\infty}^{\gamma+i\infty}\frac{s^{sw}\,ds}{|\alpha|\,\sqrt{(s-t_1)(t_2-s)}}
    \label{w7}
\end{eqnarray}
along the vertical line $Re[s]=\gamma$, lying  
in the region of convergence. Eq. (\ref{w4}) suggests that $t_1<\gamma<t_2$.

\begin{figure}[t]
  \includegraphics[width=5cm]{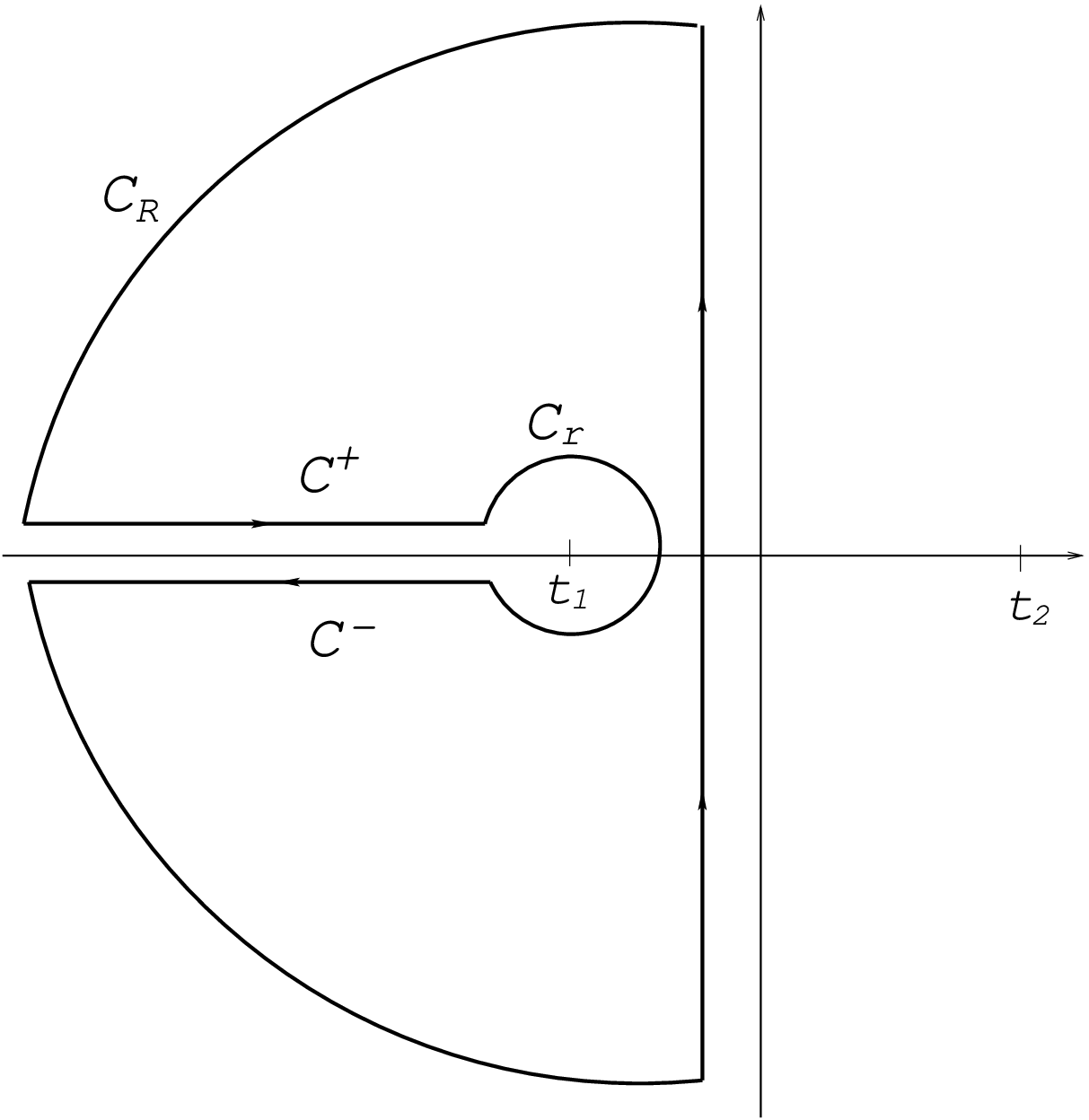}
\caption{The integration contour $\Gamma$ in Eq. (\ref{w8}).
\label{Bromwich}}
\end{figure}

For $w>0$, we close the Bromwich contour by a semi-circle to the left, bypassing the branch cut along the part of the real axis $x<t_1<0$ as shown in Fig. {\ref{Bromwich}. The resulting closed contour $\Gamma$ does not enclose any singularity, therefore 
\begin{eqnarray}
    I=\frac{1}{2\pi i}\,\int\limits_{\Gamma}\frac{s^{sw}\,ds}{|\alpha|\,\sqrt{(s-t_1)(t_2-s)}}=0.
    \label{w8}
\end{eqnarray}
The contributions from the large semi-circle $C_R$  of radius $R$ and the small circle $C_r$ about $s=t_1$ of radius $r$ vanish in the limits $R\to\infty$ and $r\to 0$, and the contribution from the vertical line equals $f(w)$. Therefore
$I=f(w)+I_C=0$, and 
\begin{eqnarray}
    f(w)=-I_{C}=-\frac{1}{2\pi i}\,\int\limits_{C}\frac{s^{sw}\,ds}{|\alpha|\,\sqrt{(s-t_1)(t_2-s)}},
    \label{w9}
\end{eqnarray}
where $C=C^++C^-$ consists of the two lines just above and below 
the branch cut. For $s\in C^+$ (above the cut), one uses the parametrization $s=t_1+\rho e^{i\pi}$, which gives
$\sqrt{(s-t_1)(t_2-s)}=i\sqrt{\rho(\Delta+\rho)}$, where
\begin{eqnarray}
    \Delta=t_2-t_1=\sqrt{4/\alpha^2+1}.
    \label{Delta}
\end{eqnarray}
For $s\in C^-$ (below the cut), $s=t_1+\rho e^{-i\pi}$, and 
$\sqrt{(s-t_1)(t_2-s)}=-i\sqrt{\rho(\Delta+\rho)}$. Taking that into account, from Eq. (\ref{w9}) we find for $w>0$
\begin{eqnarray}
    f(w)=\frac{e^{t_1 w}}{\pi\,|\alpha|}\,\int_0^\infty \frac{e^{-\rho w}\,d\rho}{\sqrt{\rho\, (\Delta+\rho)}}=\frac{1}{\pi\,|\alpha|}\,e^{w/2}\, K_0\left(
    \frac{\Delta}{2}\, w
    \right),
    \label{w_positive}
\end{eqnarray}
where 
$K_0(x)$ is the modified Bessel function of the second kind.

For $w<0$, the calculations are similar but we close the Bromwich contour by a semi-circle to the right, and the branch cut is now along the part of the real axis $x>t_2>0$.
Choosing an appropriate branch of the square-root function in Eq. (\ref{w7}), we find for $w<0$
\begin{eqnarray}
    f(w)=\frac{e^{t_2 w}}{\pi\,|\alpha|}\,\int_0^\infty \frac{e^{\rho w}\,d\rho}{\sqrt{\rho\, (\Delta+\rho)}}=\frac{1}{\pi\,|\alpha|}\,e^{w/2}\, K_0\left(
    -\frac{\Delta}{2}\, w
    \right),
    \label{w_negative}
\end{eqnarray}
Combining, for arbitrary $w$ we obtain
\begin{eqnarray}
    f(w)=\frac{1}{\pi\,|\alpha|}\,e^{w/2}\, K_0\left(
    \frac{\Delta}{2}\, |w|
    \right),
\end{eqnarray}
which is Eq. (\ref{df1}) of the main text.


\end{document}